\documentstyle[prb,aps,twocolumn]{revtex}
\def\geqap{\,\raise 2pt \hbox{$>\kern-11pt \lower 5pt \hbox{$\sim$}$}\,}
\def\leqap{\,\raise 2pt \hbox{$<\kern-10pt \lower 5pt \hbox{$\sim$}$}\,}
\begin{document}
\draft
\title{Spin and orbital ordering in double-layered manganites}
\author{Ryo Maezono and  Naoto Nagaosa}
\address{Department of Applied Physics, University of Tokyo,
Bunkyo-ku, Tokyo 113-8656, Japan}
\date{\today}
\maketitle
\begin{abstract}
\par
We study theoretically the phase diagram of the double-layered
perovskite manganites taking into account the orbital degeneracy, 
the strong Coulombic repulsion, and the coupling with the lattice 
deformation.
Observed spin structural changes as the increased doping
are explained in terms of the orbital ordering 
and the bond-length dependence of the hopping integral along 
$c$-axis.
Temperature dependence of the neutron diffraction peak corresponding to
the canting structure is also explained.
Comparison with the 3D cubic system is made.
\end{abstract}
\pacs{ 71.27.+a, 75.30.-m, 75.30.Et}
\narrowtext
\section{Introduction}
Perovskite manganites ($R_{1-x}, A_x$)$_{n+1}$Mn$_n$O$_{3n+1}$ 
(R=La, Pr, Nd, Sm ; A= Ca, Sr, Ba ; $n=1, 2, \infty$) have recently 
attracted renewed interests from the viewpoint of the close connection 
between magnetism and transport. 
\cite{chaha,helmolt,tokura95,jin94,ram,moritomo95-96}
The stability of the perovskite structure enables the preparation
of high quality single crystals with systematically changing the
carrier density and the bandwidth by controlling $R$ and $A$ atoms.
In addition, the crystal structure changes from two-dimensional
($n=1$, single layer) to three-dimensional ($n=\infty$, cubic) as
$n$ increases.
The dimensionality of the electronic structure is controlled also by
the spin and the orbital orderings, for example, the transfer along $c$-axis
is forbidden by the anti-parallel spin configuration and/or the
$d_{x^2-y^2}$ orbital ordering.\cite{maezono981,maezono982}
Therefore even in the isotropic ($n=\infty$, cubic, 113-system) 
crystal structure, the electronic dispersion can be quasi-two-dimensional 
(layered antiferromagnetic (spin $A$) in Pr$_{1-x}$Sr$_{x}$MnO$_{3}$,
\cite{kawano97,tomioka}
Nd$_{1-x}$Sr$_{x}$MnO$_{3}$,\cite{kuwahara98,kajimoto99,kuwahara99}
La$_{1-x}$Sr$_{x}$MnO$_{3}$\cite{moritomo98-113})
or quasi-one-dimensional (rod type $AF$ (spin $C$) in 
Nd$_{1-x}$Sr$_{x}$MnO$_{3}$ \cite{kuwahara98,kajimoto99}).
Another important related issue is the spin canting.
As discussed by de Gennes\cite{degenne}, $A$-type antiferromagnet
near the parent 
insulator is unstable toward the spin canting when holes are doped.
This is because the kinetic energy gain along $c$-axis wins the energy 
cost of the $AF$ exchange interaction for small canting angle.
In the metallic $A$-type $AF$ phase mentioned above, however, no spin canting
has been observed. \cite{kawano97}
\par
Double-layered manganites with $n=2$ (327-system) offer an interesting 
opportunity
to study the interplay of the spin, the orbital and the crystal structure
in the above-mentioned issues of dimensionality and spin canting.
With increasing $x$, the spin ordering within one double-layer changes
from the ferromagnetic one (spin $F$, $x=0.3$) 
to the spin $A$ ($x=0.5$).
\cite{moritomo95-96,battle96,kimura96-97,mitchell97,argyriou971,argyriou972,perring97,hirota98,moritomo97,moritomo98,kimura98,kubota98}
For $0.4<x<0.48$ the diffraction peak observed in the neutron-scattering
experiments indicates the coexistance of the spin $F$- and
$A$-components at low temperatures, which has been interpreted as 
the spin canting.
\cite{argyriou971,argyriou972,perring97,hirota98,kimura98,kubota98}
However it could be interpreted also in terms of the phase separation,
and further theoretical studies are needed.
Lattice deformation depending upon $x$ is also observed,
\cite{mitchell97,hirota98,moritomo97,moritomo98,kimura98,kubota98}
implying active contribution to the spin transition, in comparison
with 113-system.
In 113-system, there is the one to one correspondence between the crystal
deformation and the orbital ordering, e.g., the $c$-axis contracts in the
metallic spin $A$ state \cite{kuwahara99} where the $d_{x^2-y^2}$ ordering 
occurs. \cite{maezono982}
In 327-system, on the other hand, MnO$_6$ octahedra elongates along $c$-axis
during the spin structural change from spin $F$ to cant, and to spin $A$. 
\cite{moritomo98} 
This elongation decreases with increasing $x$ upto $x=0.5$ continuously, 
but even in the spin $A$ phase a little elongation remains,\cite{moritomo98} 
though this deformation prefers the $d_{3z^2-r^2}$.
\par
In this paper, we study the spin and the orbital ordering in the 
double-layered compounds.
Possibility of the spin canting is studied for general
$x$, which includes the de Gennes's canting mechanism as
a special case.
In the metallic region, the spin $A$ is locally stable against the
spin canting.
Spin transition from the spin $A$ is therefore discontinous one
to the spin canting or to the spin $F$, depending on the ratio of
the transfer integral and the superexchange interaction between
$t_{2g}$ spins.
For the spin canting to occur in the metalic region, the small transfer
integral along $c$-axis, i.e., the planer orbital 
ordering, turned out to be indispensable.
The canting angle is sensitive to the
bond-length along $c$-axis.
With this mechanism, calculated mean-field phase diagram
together with  the observed lattice distortion 
\cite{mitchell97,hirota98,moritomo97,moritomo98,kimura98}
can  qualitatively explain the observed $x$-\cite{moritomo95-96,battle96,kimura96-97,mitchell97,argyriou971,argyriou972,perring97,hirota98,moritomo97,moritomo98,kimura98,kubota98} and $T$-dependence \cite{hirota98} 
of the spin ordering 
and also the spin-anisotropy \cite{argyriou972,hirota98,kubota98,perring98}
in 327-system.
\par
The plan of this paper follows.
In Sec. II, the model and the formulation are given. 
Results and discussion are presented in Sec. III.
Notations used here are standard as in ref. 8.
%
\section{Model and formulation}
 We start with the Hamiltonian
\begin{eqnarray}
H&=&
\sum\limits_{\sigma \gamma \gamma' \langle ij \rangle} 
{t_{ij}^{\gamma \gamma '}d_{i\sigma \gamma }^{\dagger}d_{j\sigma \gamma '}}
\nonumber \\
&-&
J_H\sum\limits_i {\vec S_{t_{2g} i}\!\cdot\! \vec 
S_{e_g i}}
\nonumber \\
&+& J_S\sum\limits_{\left\langle {ij} \right\rangle } {\vec S_{t_{2g} i}
\!\cdot\! \vec S_{t_{2g} j}} +H_{\rm on\ site}
\nonumber \\
&+&H_{\rm el-ph} \ ,
\label{eqn:eq1}
\end{eqnarray}
where $\gamma$ [$=a(d_{x^2-y^2}), b(d_{3z^2-r^2})$] specifies the 
orbital and the other notations are standard.
The transfer integral $t_{ij}^{\gamma \gamma'}$ depends on the pair 
of orbitals $(\gamma, \gamma')$ and the direction of the 
bond $(i, j)$ \cite{ishihara97},
$J_H$ is the Hund coupling between $e_g$ and 
$t_{2g}$ spins, and $J_S$ is 
the $AF$ coupling between nearest neighboring $t_{2g}$ spins.
$H_{\rm on\ site}$ represents the on-site
Coulomb interactions between $e_g$ electrons, and is given by
\cite{maezono982,ishihara97}
\begin{equation}
H_{\rm on\ site}
= -\sum\limits_i {\left( {\tilde \beta \vec T_i^2+\tilde
 \alpha \vec S_{e_{g} i}^2} \right)}, 
\label{eqn : eq2}
\end{equation}
where the spin operator for the $e_g$ electron is defined as 
$\vec S_{e_g i}={1 \over 2}\sum\limits_{\gamma \alpha \beta} 
 {d_{i\gamma \alpha }^{\dagger}\vec \sigma _{\alpha \beta }
d_{i\gamma \beta }}$ with the Pauli matrices $\vec \sigma$, 
while  the orbital isospin operator is defined as 
$
\vec T_i={1 \over 2}\sum\limits_{\gamma \gamma' \sigma}  {d_{i\gamma \sigma }
^\dagger\vec \sigma _{\gamma \gamma '}d_{i\gamma '\sigma }} \ .
$
Coefficients of the spin and isospin operators, i.e., 
$\tilde \alpha $ and $\tilde \beta $, are given by 
\cite{maezono982,ishihara97}
$
\tilde \alpha = U-{J \over 2}>0\ , 
$
and 
$
\tilde \beta = U-{3J \over 2}>0 \ .
$
The minus sign in Eq. (\ref{eqn : eq2}) means 
that the Coulomb interactions induce both spin and orbital 
(isospin) moments.
 The parameters $\tilde \alpha ,\tilde \beta ,
t_0$, used in the numerical calculation are chosen as
$t_0 = 0.72$ eV, $U=6.3$ eV, and $J=1.0$ eV, being 
relevant to the actual manganese oxides. \cite{maezono981,maezono982} 
The electron-lattice interaction is \cite{maezono982}
\begin{eqnarray}
    H_{\rm el-ph}
    = +\left|g\right|r\sum_{i}{\vec v_{i}\cdot\vec T_{i}}\ ,
\label{eqn : eq3.2.16}
\end{eqnarray}
where $g$ is the coupling constant and $r$ ($\vec v_{i}$) is
the magnitude (direction) of the lattice distortion of the MnO$_6$-octahedra.
Values of $r$ and $\vec v$ are taken from the observed elongation
along $c$-axis in (La$_{1-x}$, Sr$_x$)$_3$Mn$_2$O$_7$ ($0.3<x<0.4$) 
\cite{moritomo98} leading to $r\sim0.01$, $\vec v /\!/ \hat z$.
We calculate the ground state energy by the meanfield approximation
with the two order parameters,\cite{maezono982}
\begin{eqnarray}
    \vec \varphi_S
   & = & \left\langle \vec S_{e_g} \right\rangle + {{J_H } \over 
    {2 \tilde \alpha }} \left\langle \vec S_{t_{2g}} \right\rangle \ ,\\
    \vec \varphi_T
   & = & \left\langle \vec T \right\rangle \ .
\label{eqn : eq3.2.16}
\end{eqnarray}
\par
As the exchange interaction between two double-layers
is reported to be less than 1/100 compared with the intra double-layer one, 
\cite{kimura-p}
we consider an isolated double-layer, for which the
Brillouin zone contains only two $\vec k$-points along $c$-axis.
We consider four kinds of the spin alignment in the cubic cell: spin $F$, 
$A$, $C$ and $G$ (NaCl-type).
For spin $A$, we also consider the possibility of the canting characterized
by an angle $\eta$ which is $0\ (\pi)$ for spin $F$ ($A$).
As for the orbital degrees of freedom, we consider two sublattices $I$,
and $I\!I$, on each of which the orbital is specified by the
angle $\theta_{I,I\!I}$ as \cite{maezono982}
\begin{eqnarray}
\left| {\theta _{I,I\!I}} \right\rangle =\cos {{\theta _{I,I\!I}} 
\over 2}\left| {d_{x^2-y^2}} \right\rangle +\sin {{\theta _{I,I\!I}} 
\over 2}\left| {d_{3z^2-r^2}} \right\rangle.
\label{eqn : eqN.3}    
\end{eqnarray}
We also consider four types of orbital-sublattice ordering, i.e., 
$F$-, $A$-, $C$-, $G$-type in the cubic cell.
Henceforth, we often use a notation such as spin A, orbital $G$
($\theta_I,\theta_{I\!I}$) etc..
Denoting the wave vector of the spin (orbital) ordering as
$\vec q_{S}$ ($\vec q_{T}$), the ground state energy is
given as a function of the spin ordering ($\eta$, $\vec q_{S}$),
the orbital ordering ($\theta_{I,I\!I}$, $\vec q_{T}$), and
the lattice distortion ($g$, $r$, $\vec v$).
%
\section{Results and discussions}
\subsection{Phase diagram without canting and lattice distortion}
\begin{figure}[p]
\caption{}
\label{fig : F1}
\end{figure}
\noindent
Figure. \ref{fig : F1} shows the phase diagram at zero temperature
in the plane of the doping concentration ($x$) and the $AF$ interaction
between $t_{2g}$ spins ($J_S$), without the electron-lattice coupling  
($g=0$).
Orbital shape is represented by $p$- (planer, orbital $F$ (0,0), i.e.,
$d_{x^2-y^2}$) and $n$- (non-planer, orbital G (100,-100) around $x=0.1$ 
and orbital A (-40,-40) around $x=0.7$) hereafter.
The possibility of the spin canting is not taken into account, i.e.,
$\eta=0$ or $\pi$.
For $J_S=0$, spin $F$ is the most stable for almost the whole region of $x$, 
except around $x=0$.
This ferromagnetism comes from the superexchange interaction between
$e_g$ spins for small $x\ (x\sim0.1)$, and from the double-exchange 
interaction for larger $x$ ($x\sim0.7$). \cite{maezono982,okamoto98}
Nonmonotonic behavior with double peaks ($x=0.1,\ 0.7$) of the phase 
boundary between spin $F$ and $A$ ($J_S(F\!A)$) is then attributed 
to the cross-over from the super- to the double-exchange interaction,
similarly to the case of 113-system. \cite{maezono982}
We identify the spin transition from spin $F$ to spin $A$ 
with increasing $x$  observed experimentally for $0.3<x_{exp.}<0.5$ 
\cite{moritomo95-96,battle96,kimura96-97,mitchell97,argyriou971,argyriou972,perring97,hirota98,moritomo97,moritomo98,kimura98,kubota98}
with that around $x\sim0.15$ when  $J_S\lesssim0.006$ eV
(for example, arrow (a) or (b) in Fig. \ref{fig : F1}).
This value of $J_S$ is roughly of the same order in magnitude as the 113-case. 
\cite{maezono982,maezono981}
Depicted orbital alignment is the one optimized at each point of the 
phase diagram.
For most of the phase diagram, $d_{x^2-y^2}$ is the most stable one
reflecting the two-dimensional nature of the double-layered structure.
The spin $C$ phase completely disappears.
One might regard this self-evident because of the two-dimensionality.
This, however, is not so trivial because the valency along $c$-axis
can still lower the energy by the bonding/anti-bonding splitting
for the double-layered structure.
\par
\subsection{Spin canting}
We now study the spin canting between spin $F$ and $A$.
First we consider the possibility of the spin canting with fixed
$J_S=0.004$ eV, shown as the arrow (a) in Fig. \ref{fig : F1}.
\begin{figure}[p]
\caption{}
\label{fig : F2}
\end{figure}
\noindent
The $x$-dependence of the optimized canting angle $\eta$ is 
shown in Fig. \ref{fig : F2}.
The electron-lattice coupling is not considered ($g=0$).
The canting angle changes continuously from $\pi$ (spin $A$, $x=0$) 
to zero (spin $F$, $x\sim0.1$), corresponding to the spin canting
around the parent insulator.
In this region, the orbital alignment almost remains
that for $x=0$ (orbital $G$ (100,-100)), implying that the ordering
is still dominated by the superexchange interaction in this region.
This is consistent with the interpretation that the peak of the phase boundary 
$J_S(F\!A)$ around $x=0.1$ is attributed to the superexchange interaction.
Coexisting double-exchange interaction due to the introduced carriers
competes with $J_S$, leading to the spin canting seen for $0<x<0.1$ 
in Fig. \ref{fig : F2} through the conventional mechanism 
a la de Gennes \cite{degenne} for 113-system.
Battle $et\ al$. observed the coexistance of the spin $F$ and $A$
by the powder diffraction of the poly-crystal sample 
La$_{2}$SrMn$_2$O$_7$ ($x=0$). \cite{battle97}
This can be explained by small amount of carriers introduced by the
oxygen deficiencies which brings about the spin canting by 
this mechanism for small $x$ region, as reproduced in Fig.
\ref{fig : F2}.
When $x$ goes beyond $x\sim0.125$, $\eta$ discontinuously changes from zero
to $\pi$, corresponding to the first-order transition from spin $F$ to
spin $A$.
This behavior without canting is due to the discontinous orbital transition.
With increasing the double-exchange interaction for $x>0.1$,
the orbital discontinuously changes from orbital $G$ (100,-100) to
orbital $F$ (0,0), i.e., $d_{x^2-y^2}$, in order to maximize the kinetic 
energy gain.
With fixed $J_S$ as in Fig. \ref{fig : F2}, the spin transition between 
spin $F$ and $A$
is accompanied with this orbital transition ($n$-spin $F$ to $p$-spin $A$).
Therefore, the spin canting observed for $0.4<x_{exp.}<0.5$ in the 
single-crystal
samples, \cite{argyriou971,argyriou972,perring97,hirota98,kimura98,kubota98}
therefore, cannot be reproduced in Fig. \ref{fig : F2}.
\par
In order to explain this experimental observation, we look for the possible
mechanisms for the continous change of spin/orbital structure starting
from the $p$-spin A.
One possibility is to change the orbital wave-function via the coupling
to the lattice.
In 327-system
\cite{mitchell97,hirota98,moritomo97,moritomo98,kimura98,kubota98}
the lattice constant $c$ is longer than $a$, $b$, which prefers
$d_{3z^2-r^2}$ compared with $d_{x^2-y^2}$, and $c$ changes as $x$.
This corresponds to the ``magnetic field'' to the orbital pseudo-spin,
and tends to enhance the transfer along $z$-direction and hence the spin 
canting.
\begin{figure}[p]
\caption{}
\label{fig : F4}
\end{figure}
\noindent
In Fig. \ref{fig : F4} shown the phase diagram in the plane of $x$ and
$gr$ ($g$ : coupling constant, $r$ : elongation along $c$-axis).
This $gr$ corresponds to the ``orbital magnetic field'' to prefer 
$d_{3z^2-r^2}$.
Here $J_S$ is fixed to be 0.004 eV which is comparable to the value 
appropriate for 113-system. \cite{maezono981,maezono982}
The $p$-spin $A$ state (shaded in Fig. \ref{fig : F4}(a)) is surrounded
by the discontious ``orbital spin-flop'' transition.
This is because the ``orbital magnetic field'' $gr$ is applied not 
perpendicularly but anti-parallel to the $d_{x^2-y^2}$-direction.
Therefore it seems unlikely that the spin canting occurs between the 
$p$-spin $A$ and spin $F$ with {\it different} orbital structures, e.g., 
the arrow (a) in Fig. \ref{fig : F1}.
\par
We now look for the possibility of the spin canting with the orbital
structure fixed to be $d_{x^2-y^2}$, i.e., the arrow (b) in 
Fig. \ref{fig : F1}.
The corresponding $J_S$ value might appear to be too small ($\sim$ 0.0002 eV),
but $J_S$ is very sensitive to the bond-length $l$ ( $J_S \sim l^{-14}$
as discussed later), and the lattice elongation along $c$-axis suggests 
smaller value of $J_S$ in 327-system compared with 113.
Furthermore, the $x$-dependence of $l$ induces the $x$-dependence of $J_S$
and the arrow (b) in Fig. \ref{fig : F1} has finite vertical component.
Another remark is that even though the orbital pseudo-spin $\vec T_i$
is fully polarized along $d_{x^2-y^2}$-direction, the wave-fuction is the
hybridized one with  $d_{x^2-y^2}$ and  $d_{3z^2-r^2}$.
Namely the $d_{3z^2-r^2}$ has the weight $\sim \left(t_0/\tilde \beta
\right)^2$, which induces the effective transfer $t_z$ between layers.
However $t_z$ is much smaller than $t_0$, which is crutial to the spin
canting as discussed below.
\par
\begin{figure}[p]
\caption{}
\label{fig : F5}
\end{figure}
\noindent
In Fig. \ref{fig : F5} shown the energy as a function of the angle $\eta$
between the spins of two layers for several values of $J_S$.
Spin $F$ for $J_S=0$ is due to the above-mentioned double-exchange
interaction with the weight $\sim \left(t_0/\tilde \beta
\right)^2$.
For $J_S=$ 0.42-0.5 meV, the optimized angle $\eta$ is the intermediate between
$\eta=0$ and $\eta=\pi$, and the canting state is realized.
Therefore we conclude that the spin canting occurs between $p$-spin $A$ and
$p$-spin $F$ states when $J_S$ changes.
\begin{figure}[p]
\caption{}
\label{fig : F3}
\end{figure}
\noindent
In order to study further this canting state, we introduce a simplified model
shown in Fig. \ref{fig : F3}.
We assume that the spin polarization is perfect, i.e., all the electron
spins are aligned along the ordered direction.
Therefore we consider the spinless fermions.
The density of states is taken to be a constant $N_F$ for each layer.
We take the hole-picture, and $x$ holes are occupying this constant 
density of states from the bottom.
These two bands are split into bonding and anti-bonding ones with the
splitting $\Delta=t_z\cos{\frac{\eta}{2}=t_z \xi}$.\cite{anderson}
The kinetic energy gain $\Delta E_{kin}(\xi)$ is given by 
\begin{eqnarray}
\Delta E_{kin}\left(\xi\right) 
\!\sim\!
\left\{
\begin{array}{ll}
\!\!-t_z^2\!\cdot\!N_F\cdot \xi^2 
\ \ (\rm for\ \it \xi < \xi_c \equiv \frac{x}{N_Ft_z}) 
\\
-t_z\cdot x\cdot \xi \ \ (\rm for\ \it \xi > \xi_c) 
\\
\end{array}
\right.
\ ,
\label{eqn : eq6.5.1}  
\end{eqnarray}
while the energy cost of the exchange interaction is 
$J_S\!\cos{\eta}=J_S\left(2\xi^2-1\right)$.
The lower line of Eq. (\ref{eqn : eq6.5.1}) is obtained by
de Gennes, and if this holds the canting always occurs.\cite{degenne}
The new aspect here is that $\Delta E_{kin}(\xi) \propto \xi^2$
when the splitting $\Delta=t_z \xi$ is smaller than the Fermi energy
$\epsilon_F = x/N_F$ and both the bonding and anti-bonding bands are occupied.
Therefore the spin $A$ structure ($\xi=0,\ \eta=\pi$) is at least locally
stable when $2J_S > t_z^2 N_F$.
This condition can be satisfied when the orbital is almost $d_{x^2-y^2}$
and $t_z$ is much reduced from $t_0$.
For general orbital configuration $t_z^2 N_F$ is order of magnetude
larger than $J_S$.
Now we look for the optimized $\xi$ to minimize the total energy
$\Delta E (\xi) = \Delta E_{kin}(\xi)+\Delta E_{ex}(\xi)$.
When $\xi_c > 1$ ($x > t_z N_F$), only the upper line of 
Eq. (\ref{eqn : eq6.5.1}) is relevant and $\Delta E = \left(2J_S
-t_z^2 N_F\right)\cdot \xi^2$.
Therefore $\xi$ jumps from 1 (spin $F$) to 0 (spin $A$) as $J_S$
increases across $t_z^2 N_F/2$.
\begin{figure}[p]
\caption{}
\label{fig : F6}
\end{figure}
When $\xi_c<1$, the optimized $\xi$ as a function of $J_S$ is given in 
Fig. \ref{fig : F6}.
As $J_S$ increases, the spin structure changes as spin $F$ 
($J_S < t_z x/4$) $\rightarrow$ spin canting ($t_z x/4 <J_S<t_z^2 N_F/4$)
$\rightarrow$ spin canting with fixed canting angle 
($t_z^2 N_F/4<J_S<t_z^2 N_F/2$) $\rightarrow$ spin $A$ ($t_z^2 N_F/2<J_S$).
Note that the canting angle continuously evolves from spin $F$, but jumps
at the transition to the spin $A$.
Summarizing, $x<t_z N_F$ and $t_z x/4 <J_S<t_z^2 N_F/2$ are the condition
for the occurence of the spin canting.
Considering $N_F\sim 1/t_0$, and $J_S$ is about two orders of magnitude 
smaller than $t_0$, this condition is satisfied only when $t_z$ is much
reduced from $t_0$ and $x$ is small.
Spin canting state does not appear in Fig. \ref{fig : F2} because of the large
$t_z$ for $n$-spin $F$.
\par
Spin canting observed in the metallic region $0.4<x_{exp.}<0.5$ 
\cite{argyriou971,argyriou972,perring97,hirota98,kimura98,kubota98}
implies therefore the planer orbital in this region.
This is consistent with the spin easy axis observed within the $ab$-plane
in this region \cite{argyriou972,hirota98,kubota98}
because the planer orbital leads to this anisotropy of the easy axis 
according to  the spin Hamiltonian derived by
Matsumoto \cite{matsumoto} taking into account the spin-orbit interaction.
The observed spin transition from $F$ to $A$ with increasing $x$ should 
therefore corresponds to that from $p$-spin $F$
to $p$-spin $A$ in Fig. \ref{fig : F1}, for which
the increase in $J_S$ with increasing $x$ is needed
(arrow (b) in Fig. \ref{fig : F1}).
\par
This $x$-dependence of $J_S$ can be explained in terms of the 
bond-length dependence of the inter-layer hopping integral.
According to the pseudo-potential theory \cite{harrison}, 
the hopping integral between the $d$ orbitals
depends on the bond length $l$ as $t \propto l^{-7}$,
leading to $\xi_0\propto t^z_{e_g}/\left(t^z_{t_{2g}}\right)^2
\propto l_c^{7}$,
where $l_c$ denotes the bond-length along $c$-axis.
The observed lattice contraction along $c$-axis with 
increasing $x$, \cite{moritomo97} therefore, gives rise to 
the decrease of $\xi_0$, leading to the spin structure changes
from spin $F$ to cant, and to spin $A$.
$J_S^z\propto \left(t_{t_{2g}}^z\right)^2\propto l_c^{-14}$ increases due to
this lattice contraction along $c$-axis, which can explain the 
$x$-dependence of $J_S$ needed for the change 
from $p$-spin $F$ to $p$-spin $A$, as shown by the arrow (b) 
in Fig. \ref{fig : F1}.
\par
The neutron-diffraction experiments have revealed that the samples with 
the spin canting at zero temperature ($0.4<x_{exp.}<0.48$) become
the spin $A$ at higher temperatures. \cite{hirota98}
This can also be explained in terms of the lattice deformation as follows.
The observed lattice contraction along $c$-axis 
\cite{mitchell97,kimura98,hirota98} with increasing temperature 
brings about the decrease of $\xi_0 \propto t_z/J_S\propto l^7_c$, 
leading to the spin transition toward the spin $A$ discontinously.
Hirota $et\ al.$ also attributed this $T$-dependent spin transition
to the lattice deformation. \cite{hirota98}
\par
As seen above, the observed lattice deformation plays an important
role to explain the spin transition.
In our scenario, the bond-length dependence of the inter-layer 
hopping is the most important mechanism which relates the lattice- and 
the electron-systems.
One might, however, attribute it rather to the Jahn-Teller type 
electron-lattice interaction through the electro-static coupling
(in other word, the ``orbital magnetic field'').
In this scenario, the lattice contraction along $c$-axis 
reduces the hybridization of $d_{3z^2-r^2}$
orbital as $x$ increases, namely the stabilization of $d_{x^2-y^2}$
which prefers spin $A$. \cite{hirota98}
According to this scenario, the temperature dependence of the
spin canting mentioned above is also explained by the change in the
hybridization of $d_{3z^2-r^2}$ \cite{hirota98} which is brought about by the 
temperature-depenedent lattice elongation.
This scenario, however, results in the decrease of the inter-layer
hopping as $x$ ($T$) increases, being opposite to our scenario, where 
little change in the orbital ordering and the increase of the
inter-layer hopping $t_z\!\propto l_c^{-7}$
during the spin transition are predicted.
Main distinction between these two scenarios is whether such a large 
change in the hybridization of $d_{3z^2-r^2}$ occurs during the spin transition
or not.
As seen in Fig. \ref{fig : F2} and Fig. \ref{fig : F4}, however, the spin 
transition
should be of the first order without canting if it is accompanied with the 
large change in the orbital.
Experimentally, the spin easy axis falls onto $ab$-plane
discontinously around $x_{exp}=0.32$ (spin $F$) \cite{moritomo97,kubota98} 
and the spin transition from $F$ to cant, and to spin $A$ occurs 
with this easy axis being unchanged.
This behavior implies that the hybridization of $d_{3z^2-r^2}$ decreases
discontinously before the spin transition, and during the transition
the wave-function is almost planer as $d_{x^2-y^2}$, namely the transition 
is between $p$-spin F and $p$-spin A with canting.
Though, even in this case, the ``orbital magnetic field'' brings about
a little change in the hybridization of $d_{3z^2-r^2}$ via the 
change of the weight mentioned before, 
$\left(t_0/\tilde \beta\right)^2 \rightarrow 
\left(t_0/\left(\tilde\beta-gr \right)\right)^2$,
it is a minor correction comparing with the sensitively changing
bond-length dependence, $t^z\propto l_c^{-7},\ J_S\propto l_c^{-14}$.
\par
For the spin canting, $\xi_0=t^z x/4J_S<1$, and hence
the planer orbital during the spin transition turned out to be
essential.
In 113-system, the spin $F$ phase neighboring the metallic spin $A$
has the quasi-three-dimensional orbital ordering
for the realistic parameters in the mean field theory. 
\cite{maezono981,maezono982}
Therefore the canting does not occur because the orbital structures
are {\it different} between spin $A$ and $F$, and the transition between them
is always discontinous.
In the real 113-system no anisotropy has been reported in the spin $F$
state, and there is a theoretical suggestion of the orbital
liquid state for the spin $F$ metallic region.\cite{ishihara97}
In any case the orbital state is different from the $d_{x^2-y^2}$
in the metallic spin $A$ state.
In 327-case, the layered crystal structure brings
about the two-dimensional orbital anisotropy, reflecting the interplay
of the dimensionality of the crystal structure and that of 
the orbital ordering.
\par
\begin{figure}[p]
\caption{}
\label{fig : F7}
\end{figure}
%
Figure \ref{fig : F7} summarizes our interpretation of the experiments
(notations $x_{th.}$ and $x_{exp.}$ are used to prevent 
the confusion due to the quantitative discrepancy of $x$ between the 
calculations and the experiments).
The hatched region corresponds to the spin canting phase due to the 
planer orbital for the $p$-spin $F$.
$J_S\left(FA\right)$ is the phase boundary with the 
first-order transition between spin $F$ (or canting) and spin $A$. 
Spin $F$ phase at $x_{exp.}=0.3$ \cite{kimura96-97}
corresponds to the $n$-spin $F$ phase.
With increasing $x$, spin $F$ changes from $n$- to $p$- with 
the orbital transition into $d_{x^2-y^2}$ around $x_{th.}\sim0.125$.
This transition is of the first-order, leading to the discontinous 
reorientation of the spin easy axis from being along $c$-axis 
into within the $ab$-plane by considering the spin-orbit interaction.
\cite{matsumoto}
This explains the observed spin anisotropy
\cite{argyriou972,hirota98,kubota98,perring98}
where the easy axis, pointing parallel to the $c$-axis for $x_{exp.}=0.3$,
\cite{perring98} discontinuously turns onto the
$ab$-plane at $x_{exp.}\sim 0.32$,\cite{kubota98}
for $T\lesssim100$ K.
The $p$-spin $F$ phase around $x_{th.}\gtrsim 0.125$ cants readily
within the $ab$-plane because of the decrease in $\xi_0 \propto l_c^7$ 
due to the contraction of the bond-length along $c$-axis with increasing $x$,
and then discontinously jumps to the $p$-spin $A$, corresponding to
the observation for $0.4<x_{exp.}<0.5$. 
\cite{argyriou971,argyriou972,perring97,hirota98,kimura98,kubota98}
\par
In conclusions, we have studied the spin and the orbital ordering in 
327-system, taking the orbital degrees of freedom and the 
strong Coulombic repulsion into account.
Observed $x$-dependence of the spin ordering and its anisotropy 
were qualitatively explained in terms of the orbital ordering and the
observed lattice deformation along $c$-axis.
The observed temperature depencence of the spin structure is also 
explained by the lattice deformation.
The difference between 327-system with canting and 113-system 
without canting is understood in terms of the difference in the dimensionality 
of the crystal strucuture.
\par
The authors would like to thank K. Hirota, Y. Endoh,
T. Akimoto, Y. Moritomo, S. Ishihara, W. Koshibae,
S. Maekawa, H. Yoshizawa, T. Kimira, and Y. Tokura for their valuable 
discussions.
This work was supported by Priority Areas Grants from the Ministry 
of Education, Science and Culture of Japan.
\par
%
%

%
%
\vfill 
\eject
\noindent
Figure captions
\par
\noindent
\\
Fig. \ \ref{fig : F1}. 
Phase diagram at zero temperature in the plane of the doping 
concentration ($x$) and the $AF$ interaction between $t_{2g}$ spins ($J_S$), 
without electron-lattice coupling ($g=0$).
Orbital shape is represented by $p$- (planer, orbital $F$ (0,0), i.e.,
$d_{x^2-y^2}$) and $n$-(not planer, orbital G (100,-100) around $x=0.1$ and 
orbital A (-40,-40) around $x=0.7$).
The possibility of the spin canting are not taken into account ($\eta=0,\pi$).
Depicted arrow (a) ((b)) corresponds to the spin transion with fixed
(increasing) $J_S$.
\par
\noindent
\\
Fig. \ref{fig : F2}.
The $x$-dependence of the optimized canting angle $\eta$ with $g=0$.
\par
\noindent
\\
Fig. \ref{fig : F4}.
The phase diagram as a function of the doping concentration $x$ 
and the electron-lattice coupling $g$.
$p$-spin A is always surrounded by the discontinous boundary
with respect to $\eta$.
\par
\noindent
\\
Fig. \ \ref{fig : F5}.
$\eta$-dependence of the ground state energy calculated for
a planer orbital ($\vec T$ points to the direction of $d_{x^2-y^2}$),
which saddle point corresponds to the canting angle.
\par
\noindent
\\
Fig. \ \ref{fig : F3}.
The bonding/anti-bonding splitting of the band by the 
inter-layer hopping. 
The small (large) splitting case is depicted as (a) ((b)).
\par
\noindent
\\
Fig. \ \ref{fig : F6}.
Canting saddle point $\xi$ as a function of $J_S$
(for the case $\xi_c\leq 1$).
\par
\noindent
\\
Fig. \ \ref{fig : F7}.
Schematic sketch of the calculated phase diagram and the 
correspondence with the experiments.
\par
\noindent
\\

\begin{references}
%
\bibitem{chaha}
K. Chahara, T. Ohono, M. Kasai, Y. Kanke, and Y. Kozono, Appl. Phys. 
Lett. {\bf 62}, 780 (1993).
%
\bibitem{helmolt}
R. von Helmolt, J. Wecker, B. Holzapfel, L. Schultz, and K. Samwer, 
Phys. Rev. Lett. {\bf 71}, 2331 (1993). 
%
\bibitem{tokura95}
Y. Tokura, A. Urushibara, Y. Moritomo, T. Arima, A. Asamitsu, G. Kido, 
and N. Furukawa, J. Phys. Soc. Jpn. {\bf 63}, 3931 (1994), and A. Urushibara, 
Y. Moritomo, T. Arima, A. Asamitsu,G. Kido and Y. Tokura, Phys. Rev. B 
{\bf 51}, 14103 (1995). 
%
\bibitem{jin94}
S. Jin, T. H. Tiefel, M. McCormack, R. A. Fastnacht, R. Ramesh, 
and L. H. Chen, Science, {\bf 264}, 413 (1994). 
%
\bibitem{ram}
R.A. Ram, P. Ganguly, and C.N. Rao, J. Solid State Chem. {\bf 70}, 82 (1987).
%
\bibitem{moritomo95-96}
Y. Moritomo, Y. Tomioka, A. Asamitsu, Y. Tokura, and Y. Matsui, 
Phys. Rev. B {\bf 51}, 3297 (1995); Y. Moritomo, A. Asamitsu, 
H. Kuwahara, and Y. Tokura, Nature {\bf 380}, 141 (1996).
%
\bibitem{maezono981}
 R. Maezono, S. Ishihara and N. Nagaosa, Phys. Rev. B {\bf 57}, R21822 (1998). 
%
\bibitem{maezono982}
 R. Maezono, S. Ishihara and N. Nagaosa, Phys. Rev. B {\bf 58}, 11583 (1998). 
%
\bibitem{kawano97}
H. Kawano, R. Kajimoto, H. Yoshizawa, Y. Tomioka, H. Kuwahara,
and Y. Tokura, Phys. Rev. Lett. {\bf 78}, 4253 (1997).
%
\bibitem{tomioka}
Y. Tomioka, (unpublished).
%
\bibitem{kuwahara98}
H. Kuwahara, T. Okuda,  Y. Tomioka, T. Kimura, A. Asamitsu and Y. Tokura, 
Mat. Res. Soc. Sym. Proc.  {\bf 494}, 83 (1998).
%
\bibitem{kajimoto99}
R. Kajimoto, H, Yoshizawa, H. Kawano, H. Kuwahara, Y. Tokura, K. Ohoyama,
and M. Ohashi, condmat/9902331.
%
\bibitem{kuwahara99}
H. Kuwahara, T. Okuda,  Y. Tomioka, A. Asamitsu and Y. Tokura, 
Phys. Rev. Lett. in press.
%
\bibitem{moritomo98-113}
 Y. Moritomo, T. Akimoto, A. Nakamura, K. Ohoyama, and  M. Ohashi, 
Phys. Rev. B {\bf 58}, 5544 (1998). 
%
\bibitem{degenne}
 P. G. de Gennes, Phys. Rev. {\bf 118}, 141 (1960).
%
\bibitem{battle96} 
P.D. Battle, M.A. Green, N.S. Larskey, J.E. Millburn, P.G. Radaelli, 
M.J. Rosseinsky, S.P. Sullivan, and J.F. Vente, Phys. Rev. B {\bf 54}, 
15967 (1996).
%
\bibitem{kimura96-97} 
T. Kimura, Y. Tomioka, H. Kuwahara, A. Asamitsu, M. Tamura and Y. Tokura, 
Science 274, 1698 (1996); T. Kimura, A. Asamitsu, Y. Tomioka, and Y. Tokura, 
Phys. Rev. Lett. 79, 3720 (1997).
%
\bibitem{mitchell97} 
J.F. Mitchell, D.N. Argyriou, J.D. Jorgensen, D.G. Hinks, C.D. Potter, 
and S.D. Bader, Phys. Rev. B 55, 63 (1997); J.F. Mitchell, D.N. Argyriou, 
J.D. Jorgensen, D.G. Hinks, C.D. Potter, and S.D. Bader, Mat. Res. Soc. 
Proc. 453, 343 (1997).
%
\bibitem{argyriou971} 
D.N. Argyriou, J.F. Mitchell, J.B. Goodenough, O. Chmaissem, S. Short, 
and J.D. Jorgensen, Phys. Rev. Lett. 76, 1568 (1997).
%
\bibitem{argyriou972} 
D.N. Argyriou, J.F. Mitchell, C.D. Potter, S.D. Bader, R. Kleb and 
J.D. Jorgensen, Phys. Rev. B. 55, R11965 (1997).
%
\bibitem{perring97} 
T. G. Perring, G. Appeli, Y. Moritomo, and Y. Tokura, Phys. Rev. Lett. 78, 
3197 (1997).
%
\bibitem{hirota98} 
K. Hirota, Y. Endoh, Y. Moritomo, Y. Maruyama, and A. Nakamura, 
preprint submitted to PRB Rapid; K. Hirota, Y. Moritomo, H. Fujioka, 
M. Kubota, H. Yoshizawa, and Y. Endoh, J. Phys. Soc. Jpn. 67, 3380 (1998).
%
\bibitem{moritomo97}
Y. Moritomo, Y. Maruyama, T. Akimoto, and A. Nakamura, Phys. Rev. B 56, 
R7057 (1997).
%
\bibitem{moritomo98}
Y. Moritomo, Y. Maruyama, T. Akimoto, and A. Nakamura, J. Phys. Soc. 
Jpn. 67,405 (1998).
%
\bibitem{kimura98}
T. Kimura, Y. Tomioka, A. Asamitsu, and Y. Tokura, preprint.
%
\bibitem{kubota98} 
M. Kubota, H. Fujioka, K. Ohoyama, K. Hirota, Y. Moritomo, H. Yoshizawa, 
and Y. Endoh, preprint submitted to J. Phys. Chem. Solids (1998).
%
\bibitem{perring98}
T. G. Perring, G. Appeli, T. Kimura, Y. Tokura, and M.A. Adams, preprint.
%
\bibitem{ishihara97}
S. Ishihara, M. Yamanaka, N. Nagaosa, Phys. Rev. B {\bf 56}, 686 (1997). 
%
\bibitem{kimura-p}
H.Fujioka, M.Kubota, K.Hitota, H.Yoshizawa, Y.Moritomo, and Y.Endoh,
J. Phys. Chem. Solids, in press.
%
\bibitem{okamoto98}
 S. Okamoto et al., unpublished.
%
\bibitem{battle97} 
P.D. Battle, D.E. Cox, M.A. Green, J.E. Millburn, L.E. Spring, P.G. Radaelli, 
M.J. Rosseinsky, J.F. Vente, Chem. Matter. 9, 1042 (1997).
%
\bibitem{anderson}
P. W. Anderson, and H. Hasegawa, Phys. Rev. {\bf 100}, 675 (1955).
%
\bibitem{matsumoto}
G. Matsumoto, J. Phys. Soc. Jpn. {\bf 29}, 606 (1970).
%
\bibitem{harrison}
W. A. Harrison, in {\it Electronic Structure and the Properties of Solids, 
The Physics of the Chemical Bond}, W.H.Freeman and Company, San francisco 
(1980).
%
%
%
%
%
%
%
\end{references}
\end{document}